\documentclass[11pt,a4paper]{article}
\pdfoutput=1

\usepackage{jheppub}
\usepackage{color}
\usepackage{graphicx}
\usepackage{wrapfig,enumerate,slashed}
\usepackage[utf8]{inputenc}
\usepackage{adjustbox}

\usepackage{appendix}

\usepackage{footmisc}
\usepackage{amsmath}
\usepackage{wasysym} 
\usepackage{graphicx}
\usepackage{color}
\usepackage{comment}
\usepackage{hyperref}
\usepackage{subcaption}
\usepackage{slashed}
\usepackage{booktabs}

\newcommand{\be}{\begin{eqnarray}}
\newcommand{\ee}{\end{eqnarray}}

\newcommand{\bee}{\begin{eqnarray}}
\newcommand{\eee}{\end{eqnarray}}
\newcommand{\beeq}{\begin{equation}}
\newcommand{\eeeq}{\end{equation}}

\title{Quantum Machine Learning for Particle Physics using a Variational Quantum Classifier }

\author[a,b]{Andrew Blance}
\author[a]{and Michael Spannowsky}

\affiliation[a]{IPPP, Department of Physics, Durham University, Durham DH1 3LE, UK}
\affiliation[b]{Institute for Data Science, Durham University, Durham, DH1 3LE, UK}

\emailAdd{andrew.t.blance@durham.ac.uk}
\emailAdd{michael.spannowsky@durham.ac.uk}

\abstract{
	Quantum machine learning aims to release the prowess of quantum computing to improve machine learning methods. By combining quantum computing methods with classical neural network techniques we aim to foster an increase of performance in solving classification problems. Our algorithm is designed for existing and near-term quantum devices. We propose a novel hybrid variational quantum classifier that combines the quantum gradient descent method with steepest gradient descent to optimise the parameters of the network. By applying this algorithm to a resonance search in di-top final states, we find that this method has a better learning outcome than a classical neural network or a quantum machine learning method trained with a non-quantum optimisation method. The classifiers ability to be trained on small amounts of data indicates its benefits in data-driven classification problems. 
}

\begin{document}
\preprint{IPPP/20/48}

\maketitle
\flushbottom

%%%%%%%%%%%%%%%%%%%%%%%%%%%%%%%%%%%%%%%%%%%%%%%

\section{\label{Sec:Intro}Introduction}

To discover new physics at the LHC, highly complex rare signal events have to be separated from a large number of Standard Model background events. Novel reconstruction techniques often rely on machine learning algorithms which show an outstanding ability to find correlations in high-dimensional parameter spaces to discriminate signal from background processes. In collider phenomenology, the feature space on which the machine-learning methods are trained to classify events into signal and background consists usually of physical observables of reconstructed objects, e.g. the transverse momentum of a jet $p_{T,j}$ or the total amount of missing transverse energy $\slashed{E}_T$.

The most popular machine learning techniques in recent years are artificial neural networks (NN), which are built on three pillars: 
\begin{itemize}
\item[i.] an adaptable complex system that allows approximating a complicated function, 
\item[ii.] the calculation of a loss function in the output layer which is used to define the task the NN algorithm should perform by minimising this function, and 
\item[iii.] a way to update the network continuously while minimising the loss function, e.g. through backpropagation.
\end{itemize}
In NNs the adaptable system in (i) consists of a variable number of layers made of interconnected neurons. The neurons receive inputs from previous layers in terms of weights and a bias, which are then processed as arguments of an activation function. Due to its modular setup and variable complexity an NN can be trained to perform a large spectrum of tasks\footnote{Applications range from playing Go \cite{alphago} or Chess \cite{alphazero}, over classification and image recognition \cite{imagerec, Kasieczka:2019dbj} to natural language processing \cite{language} and generative algorithms \cite{goodfellow}. Beyond classification and the regression of data points NN can also be used to find solutions to functionals and integro-differential equations \cite{Piscopo:2019txs}.}.

Quantum machine learning is an emergent research field which aims to release the prowess of quantum computing to improve machine learning methods. At the moment a full quantum neural network, where all three pillars are combined in an algorithm that is entirely built on the principles of quantum information processing, is not attainable. However, with present or at least with near-term quantum devices dedicated quantum algorithms can support pillars (i)-(iii) individually in form of a hybrid quantum machine learning approach.

In relation to NNs novel techniques are being developed and applied in a beneficial way for each of the pillars (i)-(iii) above. For example and concreteness, to support (i) quantum nodes can be connected with each other to form a variational circuit \cite{farhi2018classification, Killoran_2019, Schuld_2020} or added to a classical neural network in a hybrid approach \cite{McClean_2016, mari2019transfer}. For (ii) the loss function can be calculated using a quantum algorithm, e.g. in variational quantum algorithm approaches \cite{Peruzzo_2014, farhi2014quantum}. Whereas for (iii) one can minimise the loss function using a quantum annealer \cite{annealers_Nevan, Quant_comp_fahri} or a quantum optimisation algorithm \cite{PhysRevA.48.1687, Stokes_2020}. 

Thus, by combining quantum computing methods with a NN and applying this approach to tackle challenges in particle physics we aim to foster an increase of performance associated with quantum computing algorithms\footnote{Relevant to particle physics, a recent surge in proposals has emerged in how quantum computing can provide benefits for a variety of tasks. Quantum annealers, for example, perform continuous time quantum computations and are therefore well-suited to study the dynamics of quantum systems, even quantum field theories \cite{Abel:2020ebj,Abel:2020qzm}, and in solving optimisation problems \cite{MottQuantum}. Quantum gate computers are in particular a popular choice to calculate multi-particle processes \cite{Jordan:2011ci,Garcia-Alvarez:2014uda,Jordan:2014tma,Jordan:2017lea,Preskill:2018fag,bauer2019quantum,Moosavian:2019rxg,Alexandru:2019ozf, Alexandru:2019nsa, Lamm:2019uyc, Lamm:2020jwv}, often with field theories mapped onto a discrete quantum walk  \cite{Marque-Martin:2018PRA,Arrighi:2018PRA,Jay:2019PRA,DiMolfetta:2020QIP} or a combined hybrid classical/quantum approach \cite{Lamm:2018siq,Harmalkar:2020mpd,Wei:2019rqy,Matchev:2020wwx}.} which can then translate into an improved sensitivity in searches for novel physical phenomena. The most important task for a machine learning application in particle physics is classification. To pursue this task using quantum machine learning, we construct a novel hybrid neural network, based on a quantum variational classifier. Quantum variational classifiers are known to have an advantage in model size compared to classical neural networks \cite{Schuld_2020}. This allows us to augment the optimisation process of our hybrid network using the quantum gradient descent (QGD) method, which is inspired by the natural gradient descent method. Such complex optimisation methods are often computationally prohibitive for deep neural networks. Variational quantum classifiers are structurally very similar to classical neural networks and provide therefore an instructional framework to discuss in how far (i)-(iii) of classical NNs can be augmented using quantum computing elements.

Specifically, we use a quantum neural network approach, i.e. trainable quantum nodes connected in a circuit, and also include classic neural network elements, i.e. a bias term. During training, we use a modified quantum optimisation algorithm, based on quantum gradient descent \cite{Stokes_2020}, designed to account for the classic elements of our model.
We apply this method to a $Z'$ resonance search, which decays to a pair of top quarks \cite{Chatrchyan:2012ku, Aaboud:2018mjh, Aaboud:2019roo}. This provides a timely and realistic playground for a phenomenologically relevant classification problem. Samples of top quark pairs where one top quark decays hadronically and the other leptonically can be purified to a very high degree, i.e.~the confidence that one trains on a pure $t\bar{t}$ sample is very high. Although $t\bar{t}$ production results in jet and lepton-rich final states, for the purpose of a transparent discussion of how variational quantum classifiers can be used to support searches of new physics, we limit ourselves to only two feature variables as input to the NNs, i.e. the transverse momentum of the hardest bottom quark $p_{T, b_1}$ and the amount of missing transverse energy in the event $\slashed{E}_T$. Extending the feature space is conceptually straightforward and will improve the networks ability to discriminate between signal and background, however, it will impact on the size of the network and the number of qubits, which would prevent us from running our hybrid quantum neural network on a real quantum device.

The paper is structured as follows: Section \ref{Sec:VQC} is dedicated to a pedagogical overview to variational quantum classifiers (VQC) and to how VQC contribute to pillars (i) and (ii) in the context quantum machine learning algorithms. Section \ref{Sec:optimisation} addresses pilar (iii), where we introduce various optimisation methods applicable to the training of the NN during backpropagation. In Section \ref{Sec:Data} we outline the technical setup for the analysis on pseudo-data. Subsequently, as described in Section \ref{Sec:Training}, we train and test two different quantum machine learning models. One will use an entirely classical approach of gradient descent while the other is trained using a quantum optimisation method. To provide a baseline we compare to a classical neural network. Finally, we provide a summary and concluding remarks in Section \ref{Sec:Conclusion}. 

\section{\label{Sec:VQC} Structure of a Variational Quantum Classifier}

Variational quantum classifiers are a form of quantum neural network that can be used for supervised learning. This is achieved by designing a quantum circuit that behaves similarly to a traditional machine learning algorithm. The quantum machine learning algorithm contains a circuit which depends on a set of parameters that, through training, will be optimised to reduce the value of a loss function. This trained circuit is described in functional form by
\begin{equation}
\label{fx}
f(w, b , x) = y, 
\end{equation}
where $f$ is the network, $y$ is the network output used to calculate the loss function $L$, the network has trainable parameter $w,\ b$ and input data $x$ . Thus, the structure of a VQC shares many similarities with a traditional neural network. In both cases, the network $f$ is built from discrete modular blocks, i.e. nodes in the classical neural network while a quantum circuit is composed of quantum gates, and share techniques used for training.

\begin{figure}[ht!]
	\centering
	\includegraphics[width=0.9\textwidth]{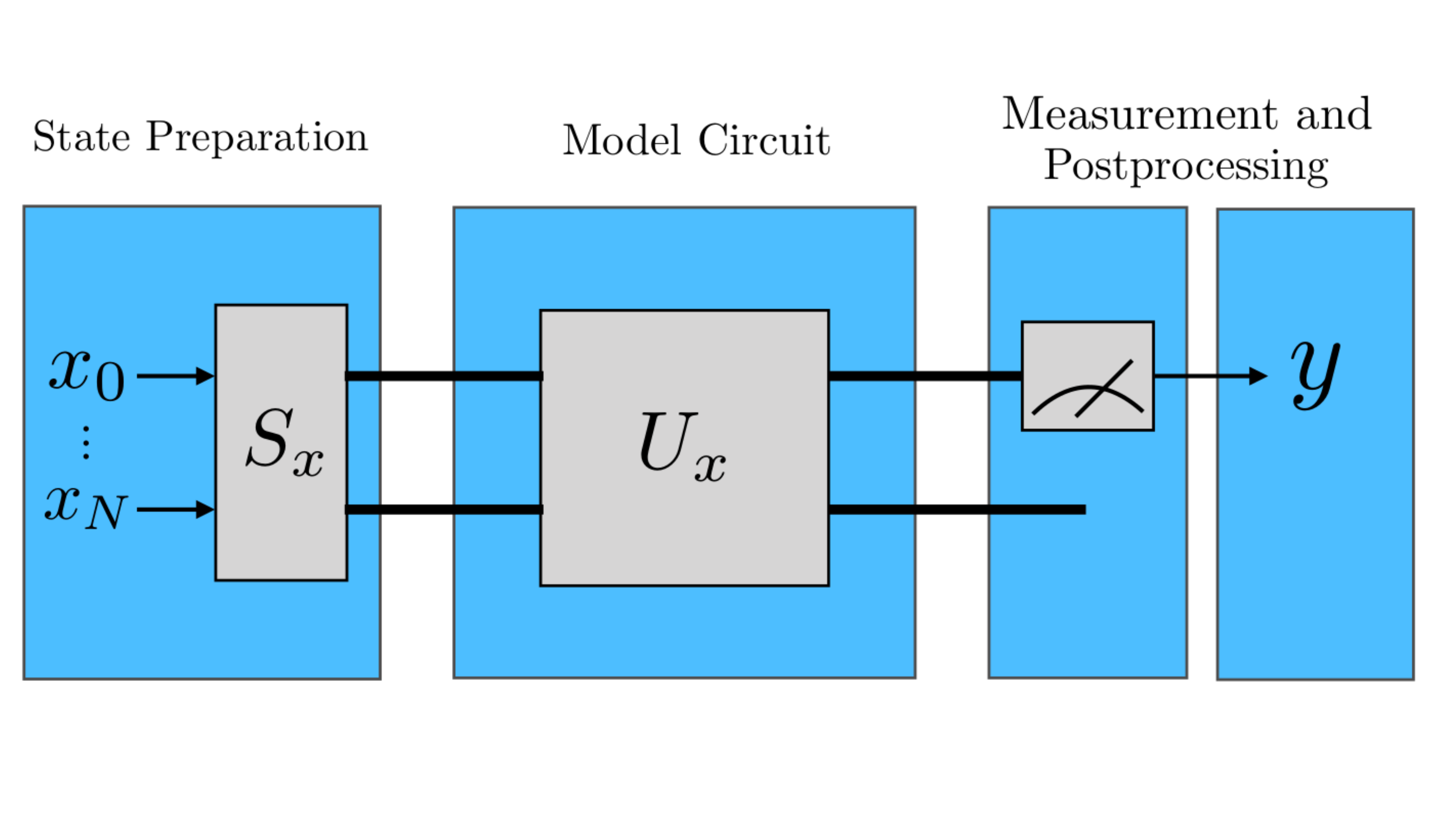}
	\caption{A variation classifier described by 3 parts. The state preparation circuit is desgined to take our input,  $x \in \mathbb{R}^{n}$, and encode it in a N-qubit quantum state. The model circuit will apply trainable and non-trainable gates to this state. In the final steps we measure the states and apply any postprocessing necessary. This model is inspired by circuit-centric quantum classifiers \cite{Schuld_2020}.}
	\label{fig:VQC_structure}
\end{figure}

Our classifier, is designed as a circuit-centric quantum classifier \cite{Schuld_2020}. It is structurally depicted in Figure \ref{fig:VQC_structure} and consists of three parts: (1) the {\it state preparation circuit}, (2) the {\it model circuit} and (3) the {\it measurement and postprocessing}.
These three parts of our model can be related in turn to the three pillars of machine learning, discussed in Section \ref{Sec:Intro}. Our classifier corresponds to (i) a complex adaptable system that (ii) calculates the value of a loss function. The continuous adaptation of the parameters $w$ and $b$, after obtaining $y$ through a measurement with the aim to continually reduce the loss function L, directly relates to the network optimisation of (iii). 

More specifically, the state preparation step, shown in Figure \ref{fig:my_circuit}, encodes the input data to an N-qubit quantum state. In classical computer algorithm this is carried out with bits, whereas on a quantum computer this is performed using qubits. A qubit is a 2-state quantum system which can be parametrised by 
\begin{equation}
\label{qubit}
\lvert \psi \rangle = \alpha \lvert 0 \rangle + \beta\lvert 1 \rangle  = \mbox{cos}\frac{\theta }{2} \lvert 0 \rangle + e^{i\varphi }\mbox{sin}\frac{\theta }{2}\lvert 1 \rangle = \binom{\mbox{cos}\frac{\theta}{2}}{\mbox{sin}\frac{\theta}{2} e^{i \phi}} \; .
\end{equation}
The state of Eq.~(\ref{qubit}) can be visualised as a vector on the Bloch sphere. By performing operations on a qubit one rotates the vector on the Bloch sphere. Circuits can be constructed to act on numerous qubits, where a 2-qubit state can be described as a tensor product of two 1-qubit states
\begin{equation}
\label{2qubit}
\lvert \psi \rangle = \alpha_{00} \lvert 00 \rangle + \alpha_{01}  \lvert 01 \rangle +\alpha_{10}  \lvert 10 \rangle + \alpha_{11} \lvert 11 \rangle \; .
\end{equation}

The model circuit is constructed from gates that evolve the input state. The circuit is based on unitary operations and depends on external parameters which will be adjusted during training. 

Finally, the postprocessing step measures the state. Traditionally, we measure the output of the first qubit.  This step will also include any classical postprocessing we may wish to include.

\subsection{State Preparation}

Before applying the model circuit of our classifier, we use a state preparation circuit $S_{x}$ to encode the input data into a quantum state. $S_x$ acts on the initial state $\lvert \phi \rangle$
\begin{equation}
\label{eq:state_prep_generic}
x \mapsto S_x  \lvert \phi \rangle = \ S_x \lvert 0 \rangle^{\otimes n} = \lvert x \rangle \; ,
\end{equation}
where $ \lvert \phi \rangle = \lvert 0 \rangle ^{\otimes n}$. The number of qubits $n$ is defined by the number of features in our dataset.

The parametrisation of the encoding can affect the decision boundaries of the classifier and can therefore be chosen in a form that suits the problem at hand \cite{larose2020robust}. Here, we use the so-called angle encoding 
\begin{equation}
\label{eq:state_prep}
\lvert x \rangle = \bigotimes_{i=1}^{n} \mbox{cos}(x_i) \lvert 0 \rangle + \mbox{sin}(x_i) \lvert 1 \rangle \; ,
\end{equation}
where $x = (x_0, ... x_N)^T$. Practically, this amounts to using the input data, $x$, as angles in a unitary quantum gate. We take the state preparation circuit as the unitary gate
\begin{equation}
\label{eq:ry}
R_y(\theta)
= \begin{pmatrix}
	\mbox{cos}(\theta/2)& \mbox{-sin}(\theta/2)\\ 
	\mbox{sin}(\theta/2) & \mbox{cos}(\theta/2)
\end{pmatrix} \; .
\end{equation}

\subsection{\label{Sec:MC}Model Circuit}

Given a prepared state, $\lvert x \rangle$, the model circuit, $U(w)$, maps $\lvert x \rangle$ to another vector $\lvert \psi \rangle = U(w) \lvert x \rangle$. In turn $U(w)$ consists of a series of unitary gates and can be decomposed as 
\begin{equation}
\label{eq:decompose}
U(w)  = U_{l_{\mathrm{max}}}(w_{l_{\mathrm{max}}})  ... U_l(w_l) ...U_1(w_1) \; ,
\end{equation}
where every $U_l(w_l)$ is a layer in the circuit, with its corresponding weight parameters, and $l_\mathrm{max}$ is the maximum number of layers. These are constructed from a set of single and two qubit gates which will evolve the state $\lvert x \rangle$. The gates include parameters that will be trained during the optimisation of the network. A single qubit gate can be written as a $2 \times 2$ unitary matrix with the form
\begin{equation}
\label{eq:generic_gate}
G(\alpha, \beta, \gamma, \phi ) = e^{i \phi} 
\begin{pmatrix}
	e^{i \beta}\mbox{cos}\alpha& e^{i \gamma}\mbox{sin}\alpha\\ 
	-e^{-i\gamma}\mbox{sin}\alpha & e^{-i\beta}\mbox{cos}\alpha
\end{pmatrix} \; .
\end{equation}
We can neglect $e^{i \phi}$ as it only gives rise to a global phase that has no measurable effect. Thus, the parameters $\alpha,\  \beta,\ $ and $\gamma$ suffice to parametrise a single qubit gate. 

We use a rotation gate, $R$, and CNOT in our model. The rotation gate is a single qubit gate that is applied to both qubits in our system. This gate is designed to rotate our state based on a set of learnable parameters $w = (\alpha,\  \beta,\  \gamma)$ \begin{equation}
\label{eq:rot_gate}
	R(\alpha,\  \beta,\  \gamma ) 
	= R_Z(\gamma)R_Y(\beta)R_Z(\alpha)
	= \begin{pmatrix}
			e^{-i (\alpha + \gamma)}\mbox{cos}(\beta/2)& -e^{-i (\alpha - \gamma)}\mbox{sin}(\beta/2)\\ 
			e^{-i(\alpha - \gamma)}\mbox{sin}(\beta/2) & e^{i(\alpha + \gamma)}\mbox{cos}(\beta/2)
		\end{pmatrix}
\end{equation}
The angles of Eq.~\ref{eq:rot_gate} are a subset of the parameters in the weight vector $w \in \mathbb{R}^{n \times 3 \times l}$, where $n$ is the number of qubits and $l$ is the number of layers in our network. This object, $w$, will contain some of the parameters that will be learned during training time. The number of qubits will mirror the number of features in our dataset whereas $l$ is a hyperparameter we can tune. In the circuit centric design we are using the number of qubits is held constant, however, the model could be extended or other frameworks used for a more flexible network design \cite{Killoran_2019}. 

\begin{figure}
	\centering
	\includegraphics[width=0.9\textwidth]{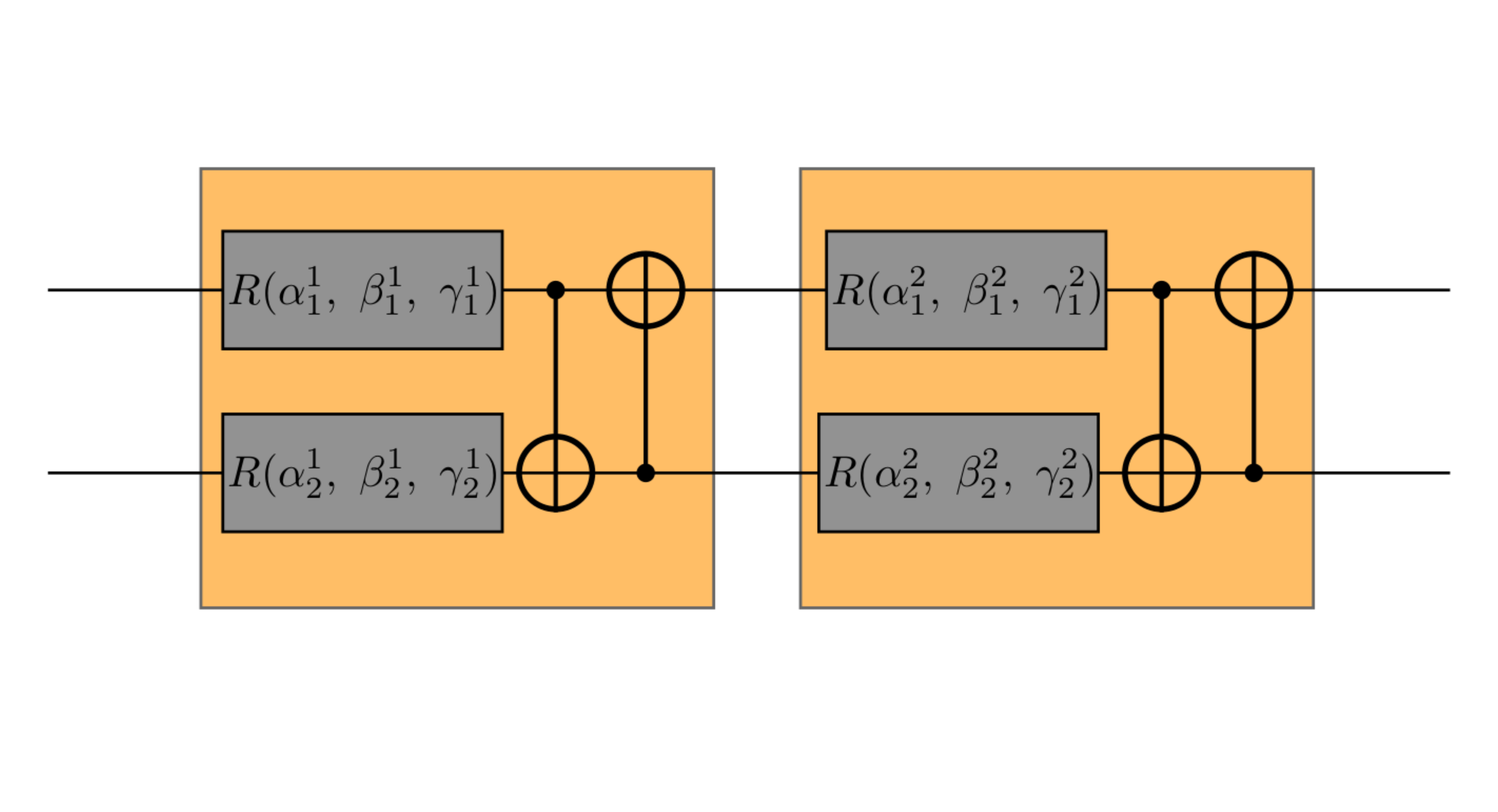}
	\caption{Circuit diagram for our variational quantum classifier model, made of two qubits in each of the two layers.}
	\label{fig:my_circuit}
\end{figure}

Each layer in our model contains two CNOT gates, a standard 2-qubit gate in quantum computing with no learnable parameters. A CNOT, if used alongside a Hadamard gate, could be used to introduce entanglement into our circuit. These gates flip the state of a qubit based on the value of another control bit\footnote{
The controlled NOT (CNOT) gate is a quantum register that can be used to entangle and disentangle quantum states. The matrix representation of a CNOT gate is
\begin{equation}
\label{eq:cnot}
\mathrm{CNOT} = \begin{pmatrix}
1 & 0 & 0 &0 \\ 
0 & 1 & 0 &0 \\ 
0 & 0 & 0 &1 \\ 
0 & 0 & 1 & 0 
\end{pmatrix}\; . \nonumber
\end{equation}}. Each gate in the layer uses a different qubit as the control bit.
The model circuit of the VQC used here is shown in Figure \ref{fig:my_circuit}.

\subsection{Measurement and Postprocessing}

After applying $U(w)$ to the initial state we need to measure its output. We do this by applying the Pauli Z operator on the first qubit and taking the expectation value 
\begin{equation}
\label{eq:expec}
\mathbb{E}(\sigma_z) = \langle 0 \lvert S_x(x)^\dagger U(w)^\dagger O U(w)S_x(x) \lvert 0 \rangle  = \pi(w, x)\; ,  
\end{equation}
where $O = \sigma_z\otimes \mathbb{I}^{\otimes(n-1)}$. To obtain an estimate we run the circuit repeatedly. The number of repetitions we do is known as the number of Shots ($S$). 

Classical postprocessing is applied to the expectation value of the circuit before returning a final classifier output. Like in a classical neural network approach, the postprocessing step gives a great deal of flexibility to the user to tackle the problem how they see fit. Generally, it will include the addition of any bias terms, the drawing of a classification decision boundary, the calculation of a loss function and the optimisation procedure.

The bias term $b$ will be a trainable parameter. Its introduction increases model flexibility and ensures the classifier output is continuous. We can write the output of our model, before thresholding, by combining the expectation value of the model circuit $\pi(w,x)$ and the bias term $b$:

\begin{equation}
\label{eq:circ_out}
f(w , b, x) =\pi(w,x) + b \; .
\end{equation}

 A decision boundary is drawn to seperate the value of $f(w,b,x)$ into the two classes. The binary classification result, $\mbox{cls}(w , b, x)$, is calculated as
\begin{equation}
\label{eq:cls}
\mbox{cls}(w , b, x) =
\begin{cases}
1 & \text{if $f(w,b,x)>0$} \;, \\
-1 & \text{else} \; .
\end{cases} 
\end{equation}

The final steps, the calculation of the loss function and carrying out the optimisation procedure will be discussed in Section \ref{Sec:optimisation}. 

\section{\label{Sec:optimisation}Optimisation}

As alluded to above, during training we aim to find the values of $w$ and $b$ to optimise a given loss function. We can perform optimisation on a quantum neural network similar to how it is done on a classical neural network. In both cases, we perform a forward pass of the model and calculate a loss function. We can then backpropagate over the network and update our trainable parameters. This is the equivalent of the third pillar of machine learning, mentioned in Section \ref{Sec:Intro}.

During training we use the mean squared error (MSE) as loss function\footnote{Often, for classification tasks using classical neural networks, the cross entropy is a preferred measure for the loss function. We find the difference between cross entropy and MSE to be irrelevant for the application discussed in Sec.~\ref{Sec:Training} and therefore follow the choice for the loss function of Refs.~\cite{iccs_variational_algorithm, Schuld_2020}}. This allows us to find a distance between our predictions and truth, represented by the value of the loss function
\begin{equation}
\label{eq:MSE}
L = \frac{1}{n}\sum_{i=1}^{n} \left (y_i^{truth} - f(w , b, x_i) \right )^2 \; .
\end{equation}
We will train our model using vanilla gradient descent and quantum gradient descent \cite{}. The latter is a quantum optimisation algorithm designed to be performed on a hybrid network we have proposed above. Further, we will exploit the advantage in model size of a variational quantum classifier compared to a classical neural network to improve its backpropagation method.

\subsection{Backpropagation }

To perform backpropagatation for a network with adjustable parameters $\theta=(w, b)$ we compute the change of it's output when varying $\theta$ as the gradient $\frac{\partial}{\partial\theta }f $. For a quantum circuits the gradient of the network output is calculated using the parameter-shift rules \cite{Mitarai_2018, Schuld_2019}. Being able to calculate gradients for quantum circuit outputs opens up the possibility of using gradient descent methods to train our variational quantum circuit. The methodology is identical to how optimisation and training is performed on classical neural networks.

For the parameter-shift rules to be correctly applied to a quantum circuit certain conditions must be met. We can represent a unitary gate in the form
\begin{equation}
\label{eq:exp}
U(\theta )= e^{i\theta V},
\end{equation}
where $\theta$ are our network parameters and $V$ is the Hermitian generator of $U(\theta )$. For a circuit $f$ that includes gates that can be represented in the form of Eq.~(\ref{eq:exp}), if $V$ has at least two distinct eigenvalues, the parameter-shift rules provide the relation
\begin{equation}
\label{eq:partial}
\frac{\partial}{\partial\theta }f = r \begin{bmatrix}
	f(\theta +s) - f(\theta -s)
\end{bmatrix} \; ,
\end{equation}
where the shift $s = \pi / 4r$. The value of $r$ is an arbitrary normalisation factor which we choose in our implementation to be $r=1/2$. Following Eq.~(\ref{eq:partial}) we can calculate gradients over quantum gates by shifting the parameters. As the difficulty of calculating $\frac{\partial}{\partial\theta }f$ has been reduced to probing the quantum circuit at different parameter points, it is now possible to evaluate the gradient fast and efficiently on a quantum device.

\subsection{From Gradient Descent to Quantum Gradient Descent }
\label{sec:qgd}

The geometry of the parameter space has a direct impact on the reliability and efficiency of an optimisation algorithm  \cite{neyshabur2015pathsgd}. Thus, a suitable choice of optimisation strategy is a key performance factor for a variational quantum circuit. It is an open question as to what is the best form of parameter space to use and whether $l_2$ Euclidean geometry is an appropriate choice for variational models  \cite{harrow2019lowdepth}.

For our problem at hand, we propose to augment the vanilla gradient descent method, often used in classical neural networks, by the quantum gradient descent method \cite{Stokes_2020}.   

In the vanilla gradient descent method, the network parameter vector $\theta_t$ at iteration step $t$ is updated with the goal that $\theta_{t+1}$ results in a smaller loss function $L(\theta)$. Thus, one approach is to update $\theta_t$ in the direction of the steepest decline, $-\triangledown L(\theta)$, weighted by the learning rate $\eta$ 
\begin{equation}
\label{eq:grad_desc}
\theta_{t+1} = \theta_{t} - \eta \triangledown L(\theta).
\end{equation}
However, this optimisation is performed on the geometry of an $l_2$ vector space, which influences the performance and how new parameters are found. While all parameters are updated with the same step size, the rate at which the loss function changes for each model parameter can vary greatly. By using this form of gradient descent it is possible to miss the global minimum in the space of the loss function. An improvement would be to change the coordinate system to ensure the loss function changed consistently with each step for each parameter or to find an optimisation method that was invariant under re-parametrisation. 

One way to address this problem is the use of natural gradient descent, which makes use of the Fisher Information Matrix \cite{Amari_1998, amari2018fisher} and is a classical extension to vanilla gradient descent method. The parameters of a network (the weights and biases) exist on a parameter space that has a Riemannian geometry. The Fisher Information Matrix is the metric that defines this space. Since this metric includes information on the geometric structure of the Riemannian space of the network parameters, its inclusion into the gradient descent optimisation leads the network to learn more effectively. 
In addition, it is invariant under re-parametrisation, and thus advantageous in finding an effective parametrisation. Algorithmically natural gradient descent can be written as
\begin{equation}
\label{eq:nat_grad_desc}
\theta_{t+1} = \theta_{t} - \eta F^{-1}\triangledown L(\theta) \; ,
\end{equation}
where $F$ is the Fisher Information Matrix. In each optimisation step, the parameters are updated in the direction of steepest descent of the information geometry rather than the Euclidean $l_2$ geometry. Although the inclusion of $F^{-1}$ in Eq.~(\ref{eq:nat_grad_desc}) in general improves the performance of the optimisation algorithm, in most classical deep neural networks calculating the inverse of a large matrix becomes computationally prohibitively expensive. However, in our hybrid network, which benefits from a small model size, the parameter space is rather small. Thus, our aim is to use a quantum optimisation equivalent of this method that we can use on variational circuits. 

The parameter space of quantum states does indeed have a geometry that can be described by an invariant metric. Similar to how the Fisher Information Matrix is used to promote the gradient descent method to the natural gradient descent method, the Fubiny-Study metric $g$, derived and elaborated on in Appendix~\ref{appendix:fubini}, exploits the geometric structure  of the variational quantum classifier's parameter space to establish the quantum gradient descent method. Here, the optimisation algorithm reads \cite{Stokes_2020}
\begin{equation}
\label{eq:quant_grad_desc}
\theta_{t+1} = \theta_{t} - \eta g^{+}\triangledown L(\theta) \; ,
\end{equation}
where $g^+$ is the pseudo-inverse of the Fubini-Study metric. We implement this algorithm using the \textsc{PennyLane} package \cite{bergholm2018pennylane}, which will  allow us to find the steepest descent in the parameter space of the quantum states. 
The approach of Eq.~(\ref{eq:quant_grad_desc}) is designed to optimise the parameters of the quantum variational circuit only, i.e. the quantum gates with trainable parameters $w=(\alpha, \beta, \gamma)$. To perform a full optimisation of our hybrid model we need to consider the classical components of our model - the bias. Thus, we propose to optimise our weights using quantum gradient descent (\ref{eq:quant_grad_desc_w}) while using vanilla gradient descent for the classical bias term $b$. Calculating both gradients at each optimisation step,
\begin{eqnarray}
\label{eq:quant_grad_desc_w}
\theta_{t+1}^{w} &=& \theta_{t}^{w} - \eta g^{+}\triangledown^{w} L(\theta) \;, \nonumber \\
\theta_{t+1}^{b} &=& \theta_{t}^{b} -\eta \triangledown^{b} L(\theta) \;,
\end{eqnarray}
 ensures our entire range of parameters is optimised simultaneously.

\section{\label{Sec:Data}Analysis Setup}
The background and signal samples used here consist of $pp \rightarrow t\bar{t}$ events and $pp \rightarrow Z^\prime \rightarrow t\bar{t}$ events, respectively. The background events have been generated with a centre-of-mass energy of $14$~TeV. When the top quarks are decayed we have forced one quark to have a hadronic decay while the other has a leptonic decay. A heavy new boson, $Z^\prime$ \cite{Altarelli:1989ff}, is used as signal, with a mass of $2$~TeV and a width chosen to be $89.6$~GeV \cite{Blance:2019ibf}. Similar to the background one top quark decays hadronically and the other leptonically. For all events, a cut of $p_T > 500$~GeV is placed on the transverse momentum of the top quarks. All events are generated using \textsc{MadGraph5\_{}aMC@NLO} \cite{Alwall:2014hca} while the parton showering and hadronisation is performed with \textsc{Pythia 8.2}.

Using the Cambridge-Aachen algorithm \cite{Dokshitzer:1997in} the hadrons and the non-isolated leptons are clustered into jets with radius $R=1.0$. This is based on work using fat jets to reconstruct highly boosted top quarks \cite{Plehn:2011tg,Plehn:2011sj,Soper:2012pb,Soper:2014rya}. Using \textsc{FastJet} \cite{Cacciari:2011ma} the $k_T$ algorithm is implemented to recluster the hardest two fat jets into jets with radius $R=0.2$. Based on proximity to a $B$-meson, jets are $b$-tagged while requiring them to have a transverse momentum $p_T > 30$~GeV. We also demand any isolated leptons to have a transverse momentum $p_T > 10$~GeV. 

The selection of these events is then based on numerous criteria. For the two fat jets in an event, one must contain at least one $b$-jet while the other must contain at least two light jets and one $b$-jet. The events must also contain a minimum of one isolated lepton and are required to have a scalar-summed transverse momentum of $H_T>1$~TeV.

\begin{figure}[!t]
	\centering
	\begin{subfigure}{0.49\linewidth} \centering
		\includegraphics[width=\textwidth]{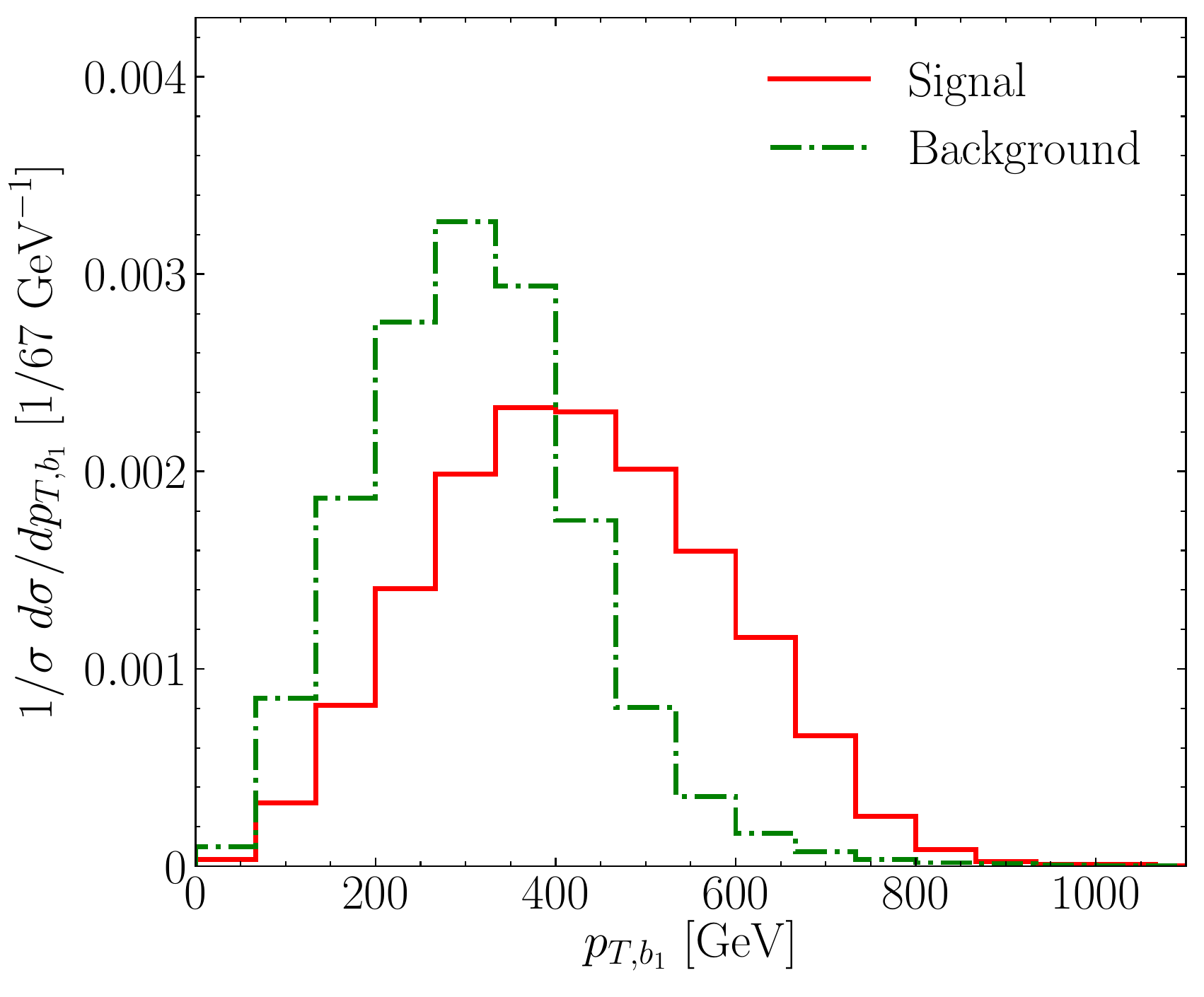}
		\caption{}
	\end{subfigure}
	\begin{subfigure}{0.49\linewidth} \centering
		\includegraphics[width=\textwidth]{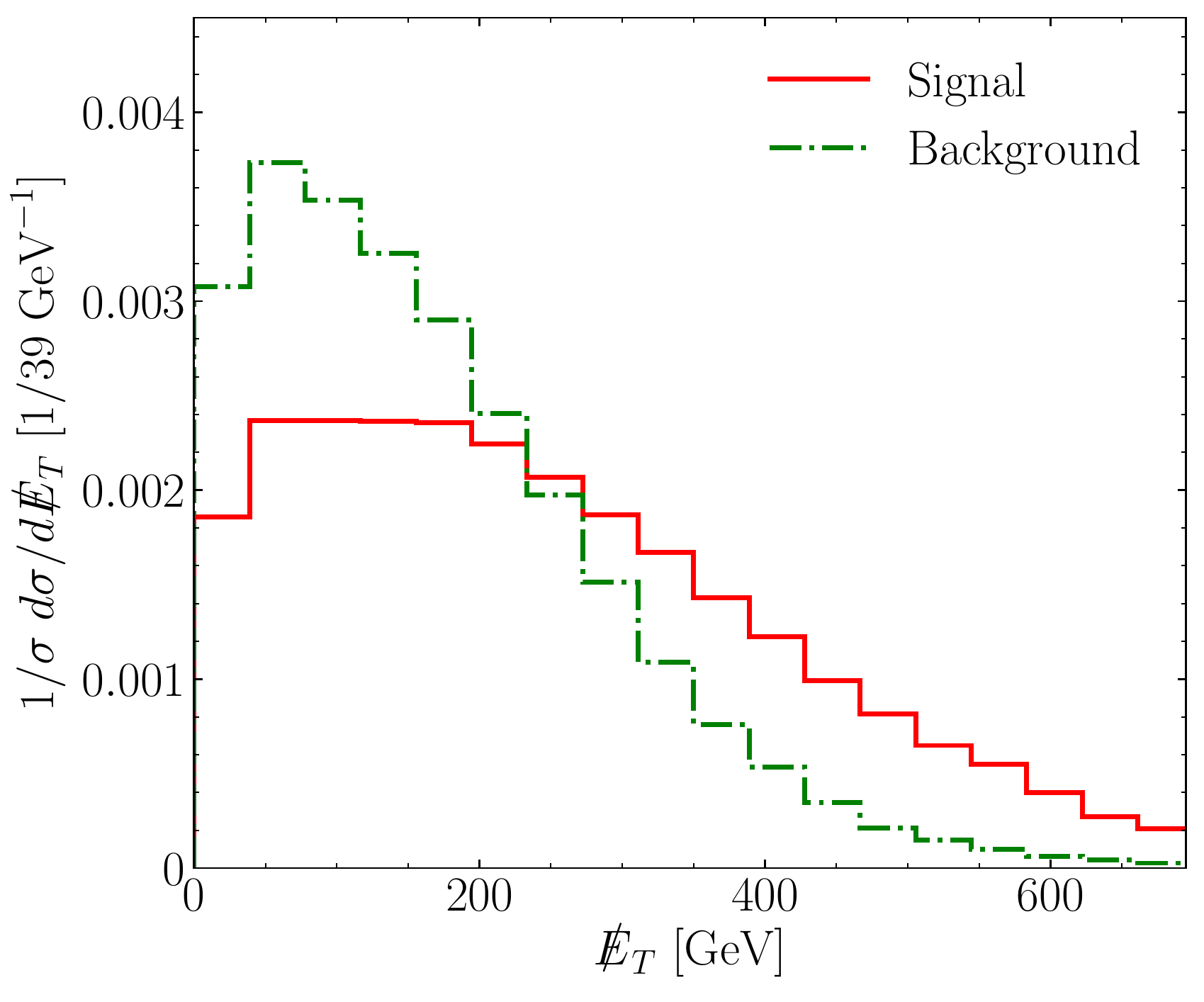}
		\caption{}
	\end{subfigure}
	\caption{Distribution of signal and background of the (a) $p_T$ of the hardest $b$-jet and the (b) missing energy.}
	\label{fig:observables}
\end{figure}

\begin{figure}[!t]
	\centering
	\begin{subfigure}{0.49\linewidth} \centering
		\includegraphics[width=\textwidth]{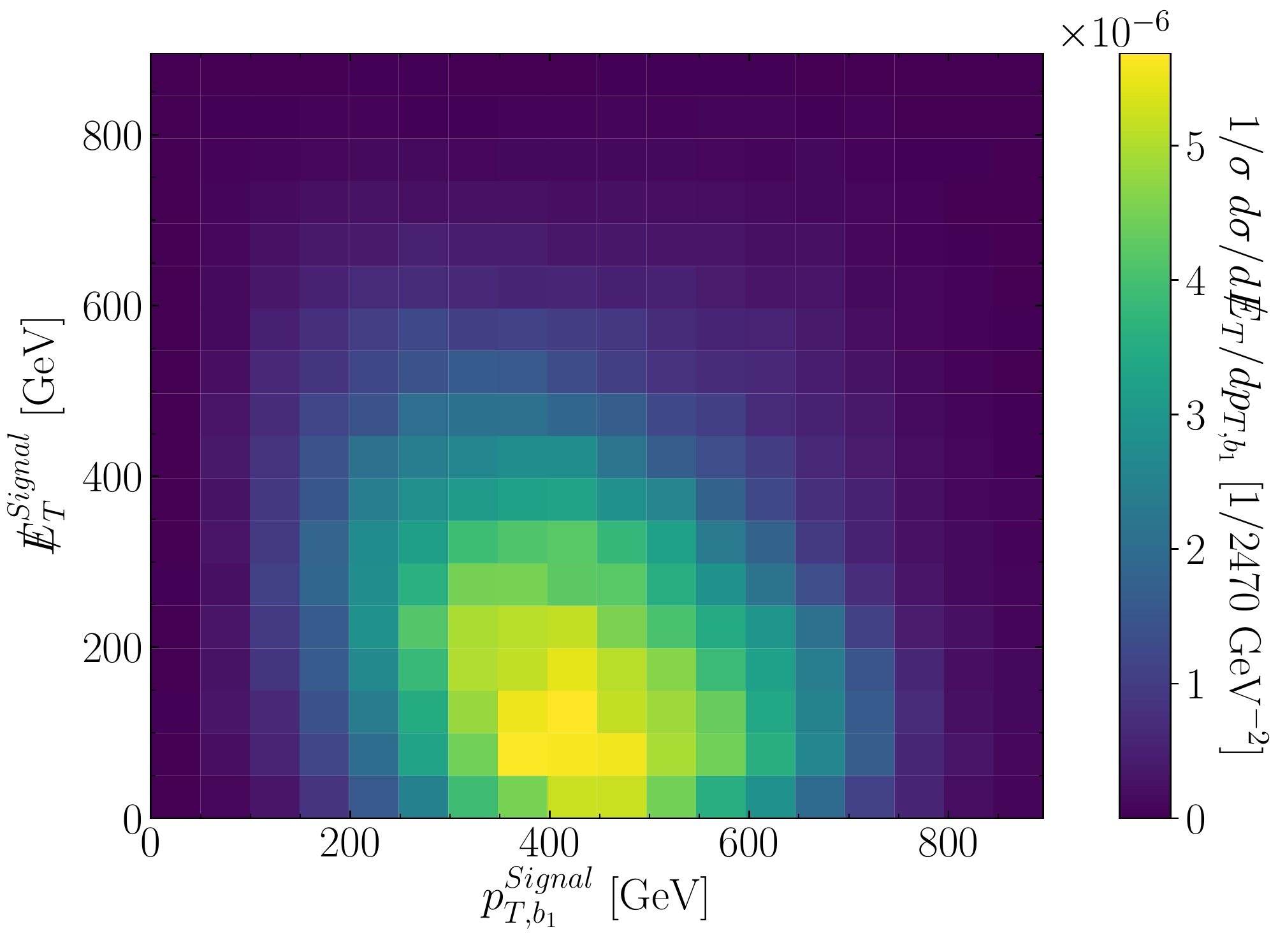}
		\caption{}
	\end{subfigure}
	\begin{subfigure}{0.49\linewidth} \centering
		\includegraphics[width=\textwidth]{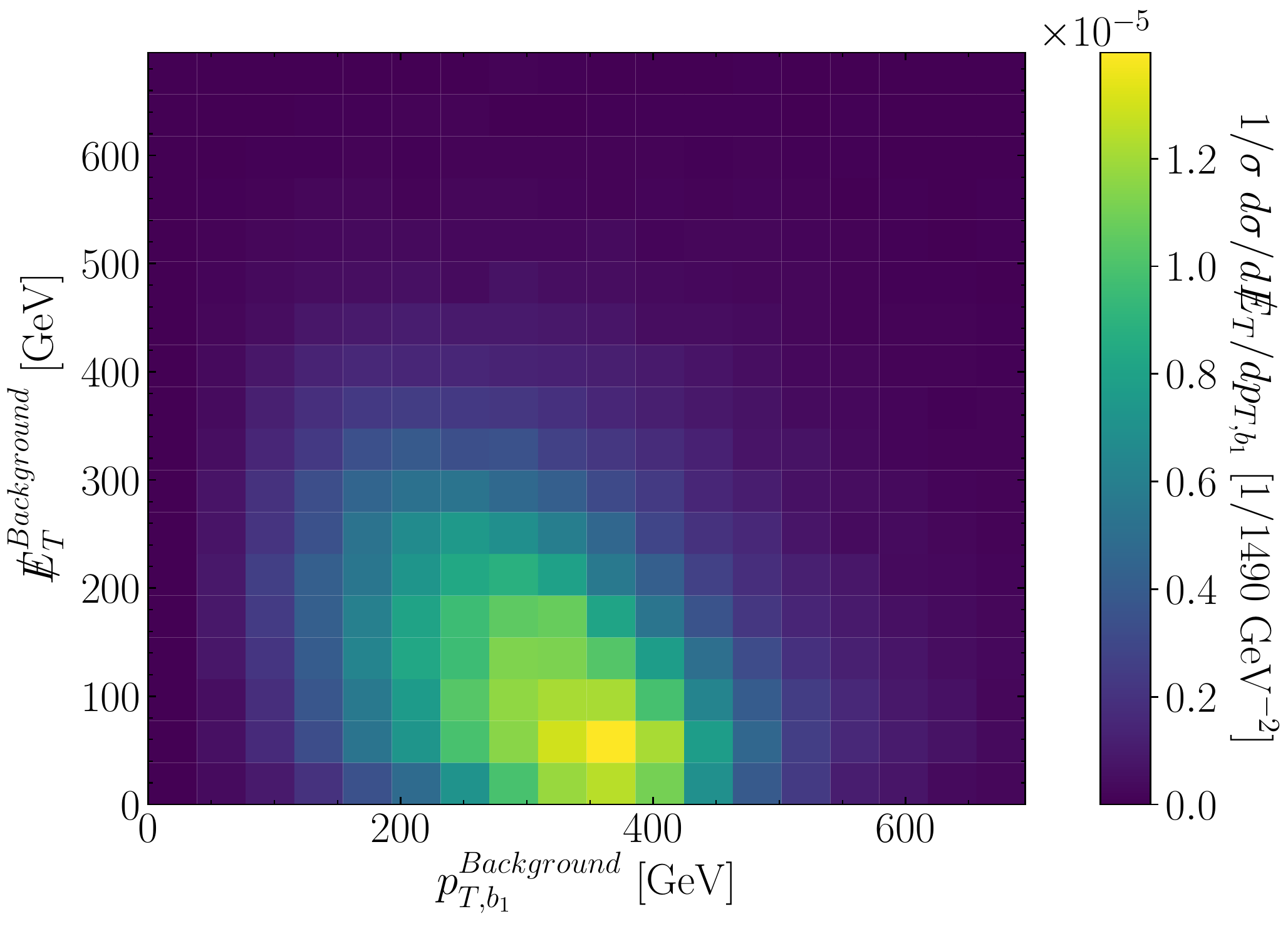}
		\caption{}
	\end{subfigure}
	\caption{Heatmaps of signal and background of the (a) $p_T$ of the hardest $b$-jet and the (b) missing energy.}
	\label{fig:obs_heatmaps}
\end{figure}

In the following, the analysis performed is exclusively based on the transverse momentum of one $b$-jet ($p_{T, b_{1}}$) and the event's missing energy ($\slashed{E}_T$). We show these observable's distributions in Figure \ref{fig:observables} and heatmaps in Figure \ref{fig:obs_heatmaps}. 

Our data $x$ is normalised using min-max scaling such that $x_{scaled} \in [0, \pi]$. This allows our features to be encoded as an angle in a qubit rotation when we begin training. The target labels are defined as $-1$ for the background set and $1$ for the signal set.

\section{\label{Sec:Training}Network Performance}

We are comparing three models: a classic neural network trained with standard gradient descent (NN-GD), a VQC trained with standard gradient descent (VQC-GD) and a VQC trained with our quantum gradient descent method (VQC-QGD) of Sec.~\ref{sec:qgd}. 

The VQC model consists of two qubits, corresponding to the two features $p_{T, b_{1}}$ and $\slashed{E}_T$, and two layers. Each layer has a rotation gate for each qubit followed by two CNOT gates. We implement this model, depicted in Fig.~\ref{fig:my_circuit}, using \textsc{PennyLane}  \cite{bergholm2018pennylane} and train it for 30 epochs with a batch size of 32 events and an initial learning rate of $\eta= 0.01$. During training, for all models, we reduce the learning rate value whenever the loss plateaus. However, learning rate reduction, in this instance, appears to have little effect on the performance of the network during training. The networks poor capacity to discriminate signal from background is reflective of the similarity between the two. Figure~\ref{fig:obs_heatmaps} shows the probability density for the events to populate areas in the feature space ($p_T$, $\slashed{E}_T$). The similarity between signal and background prevents the networks to benefit from a continuous learning rate reduction, for classical NNs and our hybrid method alike.

We anticipate that a significant advantage of the variational quantum classifier lies in its smaller network structure, which allows to employ computationally more expensive optimisation algorithms, as detailed in Sec.~\ref{Sec:optimisation}, giving in turn rise to a faster learning rate. Such a method would be particularly advantageous in cases where one has to train directly on a limited amount of data, e.g. rare decays or processes with small production cross section. 

Thus, to compare the network's ability to learn quickly, we limit ourselves to a total of 2500 events for the signal and background samples respectively. We impose a 60-20-20 split between training-validation-test sets, i.e. we train on 1500 events. To get an understanding of the effect the size of training samples have on the model performance, we train a second set of models using only 500 events each. While we carry out the training on the PennyLane's inbuilt simulator throughout, we test their performance on the PennyLane simulator, the IBM Q simulator\footnote{32-qubit backend: IBM Q team, "IBM Q simulator backend specification V0.1.547," (2020). Retrieved from https://quantum-computing.ibm.com} and IBM Q Yorktown\footnote{5-qubit backend: IBM Q team, "IBM Q 5 Yorktown (ibmqx2) backend specification V2.1.0," (2020). Retrieved from https://quantum-computing.ibm.com}. Accessing the IBM hardware was done through PennyLane's Qiskit plugin \cite{mckay2018qiskit, Qiskit}. For all backends, in training and testing, we use a total of 8192 shots.

To provide a baseline we trained a classical neural network with a vanilla gradient descent optimiser. To provide a fair and instructive comparison the network has been constructed to have a similar number of trainable parameters as the variational classifier model. The network consists of one hidden layer with 3 nodes and a ReLu activation function. The rest of the hyperparameters match what was used to train the variational classifier. To implement the network we used Keras \cite{chollet2015keras} with a TensorFlow backend \cite{abadi2016tensorflow}. 

\begin{figure}[!t]
	\centering
	\begin{subfigure}{0.49\linewidth} \centering
		\includegraphics[width=\textwidth]{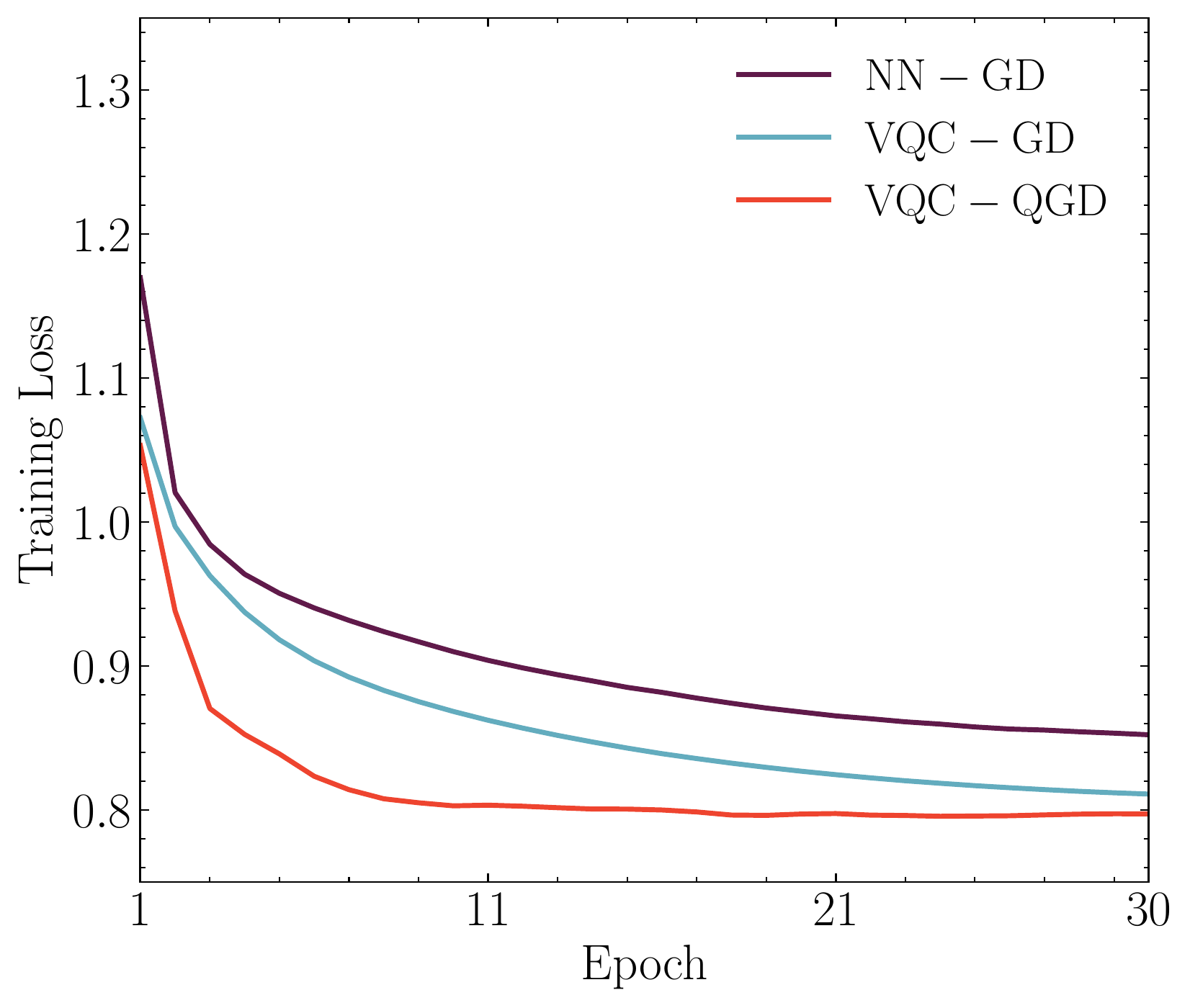}
		\caption{}
	\end{subfigure}
	\begin{subfigure}{0.49\linewidth} \centering
		\includegraphics[width=\textwidth]{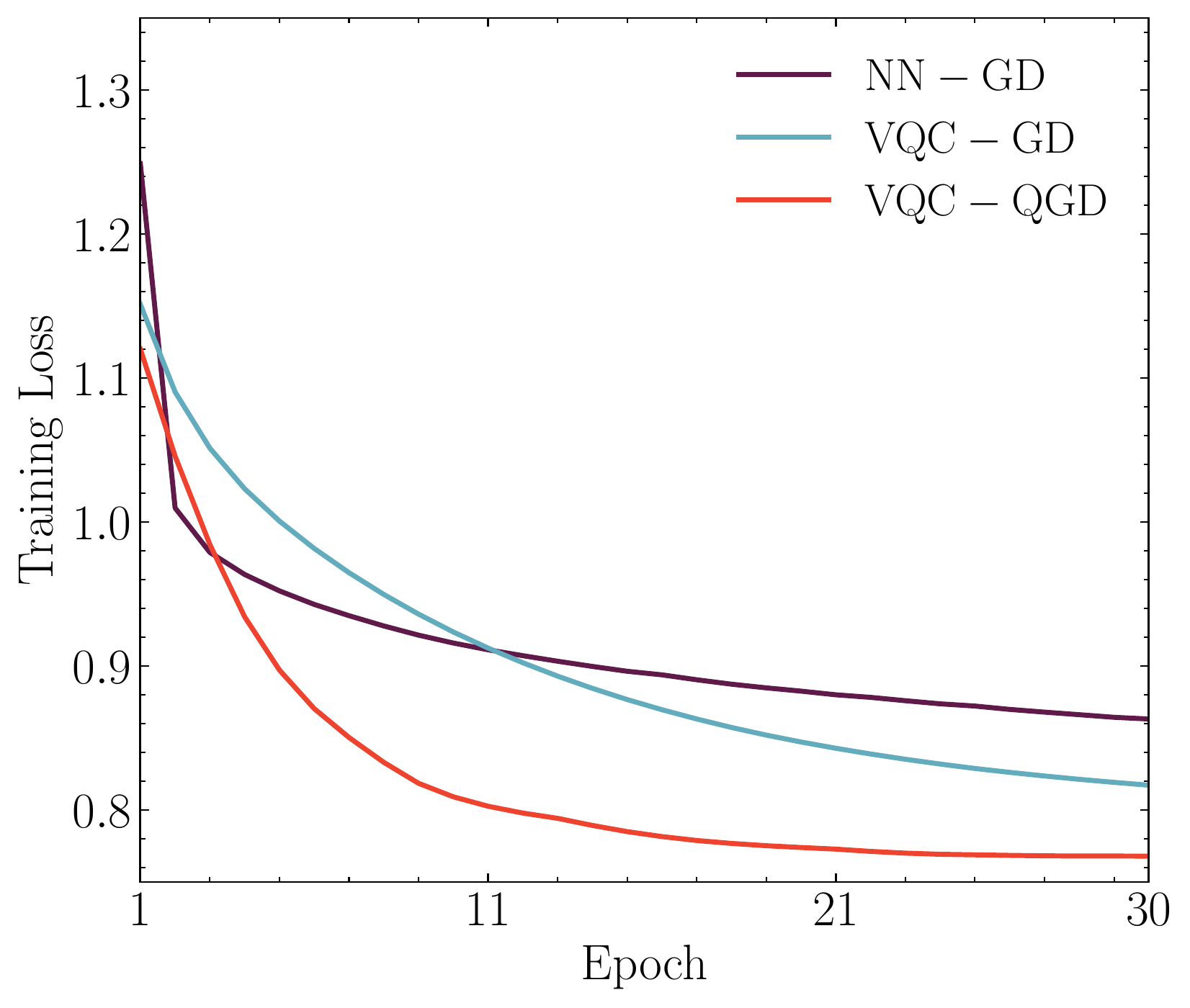}
		\caption{}
	\end{subfigure}
	\caption{Comparison of the averaged training history for 15 runs of the QVC models trained with quantum gradient descent, QVC models trained using vanilla gradient descent and the classical NN models. Figure (a) show models trained with 1500 samples and Figure (b) shows models trained with 500 samples.}
	\label{fig:learning_comparison}
\end{figure}

We found that training a classical network of this size was unstable, sometimes resulting in the loss plateauing around 1 and being unable to classify the samples. To account for the instability we saw during training of the classifier we ran each model 15 times. The results presented in Figure \ref{fig:learning_comparison} show the average loss from these runs, for each model. We see, from Figure \ref{fig:learning_comparison}, optimisation using the quantum gradient leads to a faster convergence than using the traditional gradient descent optimisation and the classical neural network. 

\begin{figure}[!t]
	\centering
	\begin{subfigure}{0.49\linewidth} \centering
		\includegraphics[width=\textwidth]{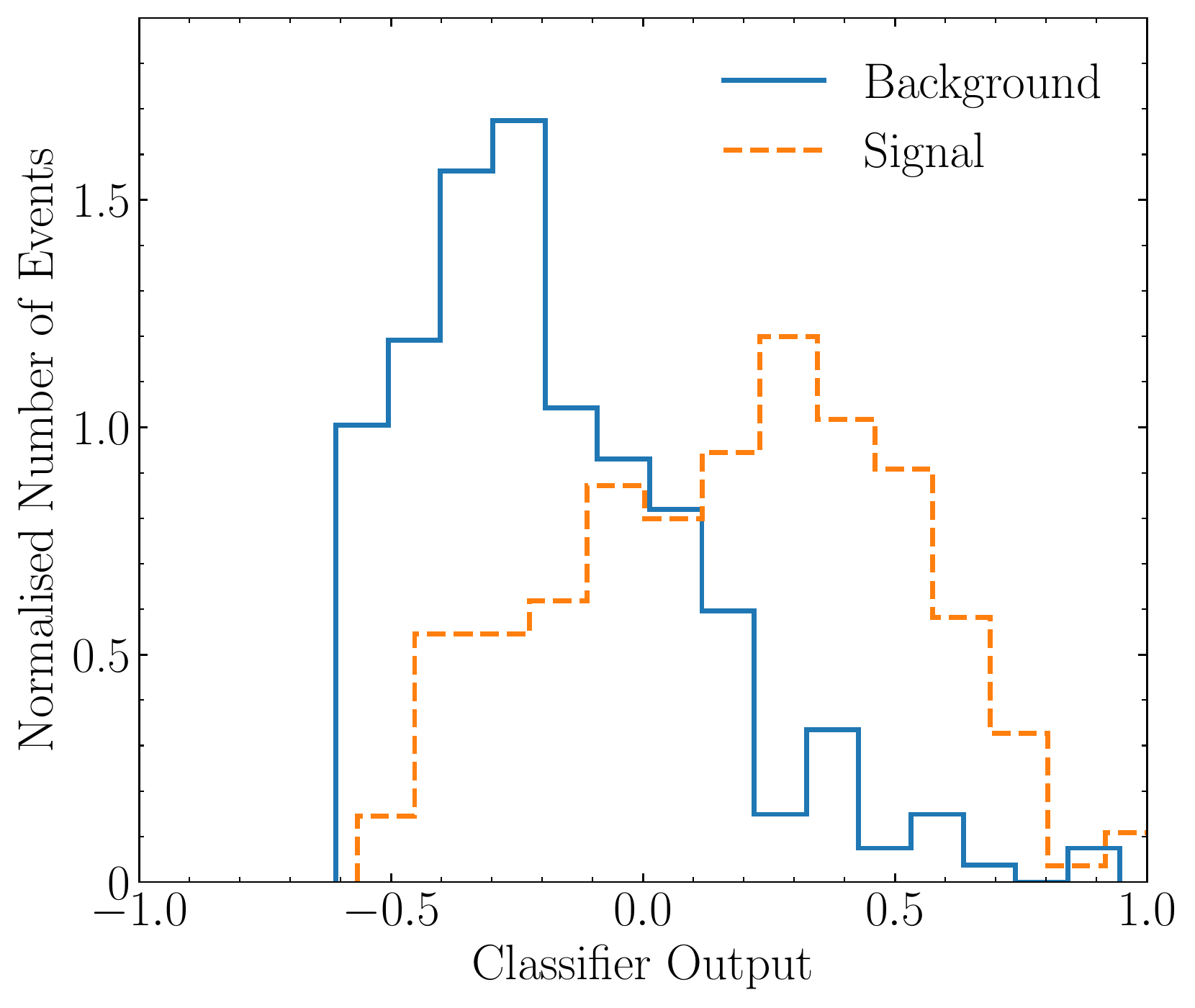}
		\caption{}
	\end{subfigure}
	\begin{subfigure}{0.49\linewidth} \centering
		\includegraphics[width=\textwidth]{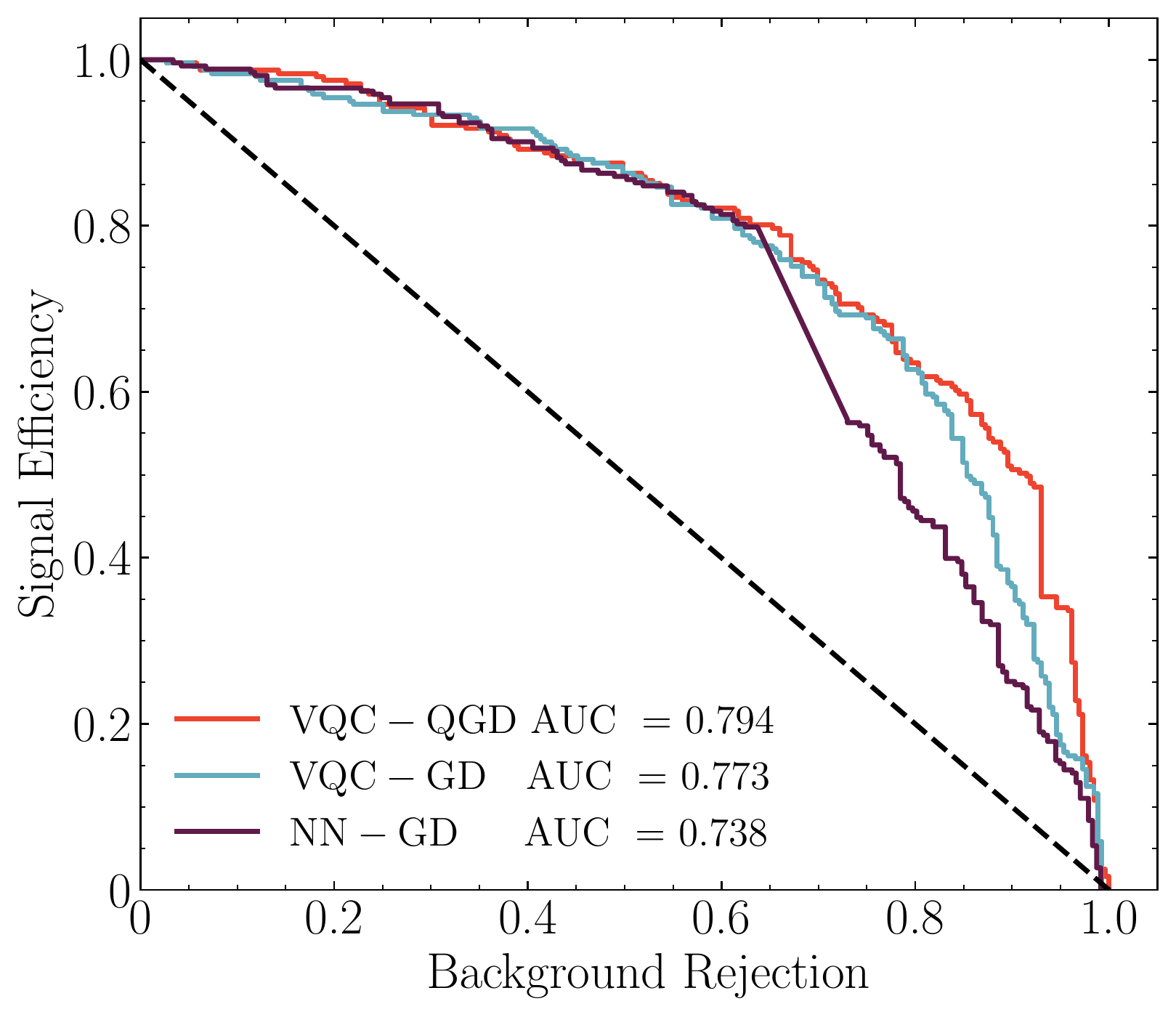}
		\caption{}
	\end{subfigure}
	\caption{ (a) Output of a QVC model trained with quantum gradient descent and (b) ROC curve for a QVC model trained with quantum gradient descent, a QVC model trained with vanilla gradient descent and the classical NN.}
	\label{fig:test}
\end{figure}

Out of each of the three sets of 15 trained models, one was chosen that had a loss value that had converged to a point during training that was similar to the average. These models where used for testing. Figure \ref{fig:test} shows the ROC curve for the chosen VQC-QGC, VQC-GD and NN-GD models. Table \ref{tab:test_results} shows the performance of the quantum gradient descent method when the test data is applied to it. We see that the model, trained on the simulator, still performs well on the real hardware. In Figure \ref{fig:test} we see an example of the variational classifier output before the decision boundary is applied and the ROC curve for each model.  

\begin{table}[ht]
	\centering
	\begin{tabular}[t]{lcc}
		\hline
		\textbf{Device}&\textbf{Accuracy (\%)}\\
		\hline
		PennyLane default.qubit	&  72.6\\ %0.766
		ibmq\_qasm\_simulator	& 72.6\\ %0.764
		ibmqx2	&71.4 \\
		\hline
	\end{tabular}
	\caption{Test set results from model trained with quantum gradient descent sent to PennyLanes in-built simulator, IBM Q simulator and IBM Q Yorktown (ibmqx2).}
	\label{tab:test_results}
\end{table}%

\section{\label{Sec:Conclusion}Conclusions}

One of the tasks with paramount importance for searches of new physics at collider experiments is the design of methods to distinguish rare signal events from large Standard Model backgrounds. In recent years increasing effort was dedicated to developing novel machine learning methods to help find correlations in high-dimensional parameter spaces. Harnessing the advantages found in quantum computing and combining them with classical neural networks to form a hybrid approach would provide another way to continue the improvement of these algorithms, possibly already accessible on near-term devices.

Quantum machine learning is an emergent research field that aims to apply these benefits to machine learning. To explore the potential quantum advantage that could come along with quantum machine learning we propose a novel hybrid neural network, based on a variational quantum classifier. Variational quantum classifier models are in many ways analogous classical neural networks. An advantage that a VQC classifier provides over a classical neural network is its small model size. The model proposed uses a quantum algorithm equivalent of natural gradient descent. Typically, due to the need to invert large matrices, natural gradient descent is computationally prohibitive on deep neural networks. However, thanks to the model-size advantage of the VQC we can make use of quantum gradient descent to optimise our network.

Thus, we combine the use of quantum gradient descent to optimise the quantum gate parameters in the model while using classical gradient descent to optimise the classical bias term. This model was used to perform a $Z^\prime$ resonance search. We compared the performance against a purely classical neural network and a VQC optimised with standard gradient descent. The hybrid approach proved successful in maximising the learning outcome. The hybrid approach learns faster than an equivalent classical neural network or the classically trained VQC. Even on small data samples the hybrid VQC still retains a high classification ability. While we applied this methodology to generated data we believe this approach can prove useful in data-driven classification problems where there is a small amount of data available.

\vskip 1 \baselineskip 
\vspace{1.0cm}
\noindent {{\it Acknowledgements.} We acknowledge use of the IBM Q for this work. We thank Steve Abel for helpful discussions. We acknowledge funding from the STFC under grant ST/P001246/1 and ST/T001011/1.}

\newpage

\appendix

\section{The Fubini-Study metric and the Quantum Geometric Tensor}
\label{appendix:fubini}

Geometric quantum mechanics states that the traits of a quantum system can be described by geometric features on a complex projective space. In this space, there is an invariant metric tensor, the Fubini-Study tensor (FST), that can be used to describe distances between quantum states \cite{kibble1979, Brody_2001, cheng2013quantum}. The FST can be found by taking the real part of the Quantum Geometric Tensor (QGT). 

We will give a general introduction to this tensor, before briefly discussing how it can be approximated on real hardware and how it relates to our VQC. We will construct the QGT by investigating the distance between the two states $\left | \psi _0(\theta ) \right \rangle$ and $\left | \psi _0(\theta + d \theta ) \right \rangle$, where $\psi$ is a general wave function state.
We can write the probability to excite the parameter from $\theta$ to $\theta +  d \theta$ as 
\begin{equation}
\label{eq:distance}
ds^2 \equiv 1-   \left | \left \langle \psi _0(\theta) |\psi _0(\theta + d \theta ) \right \rangle \right |^2     \; .
\end{equation}

The amplitude of a state being excited from $\left | \psi _0(\theta ) \right \rangle$  to $\left | \psi _n(\theta ) \right \rangle$ can be written as
\begin{equation}
\label{eq:excite}
a_n =   \left \langle  \psi _n(\theta + d \theta )| \psi _0(\theta ) \right \rangle   \; ,
\end{equation}
whereas the probability for a transition between states to occur can be found by evaluating
\begin{equation}
\label{eq:trans}
ds^2 = \sum_{n\neq 0} \left | a_n^2  \right | = d\theta_i d \theta_j  G_{i j}+ O(|d \theta|^3) \; ,
\end{equation}
where $G_{i j}$ is the Quantum Geometric Tensor, defined as
\begin{equation}
\label{eq:qgt}
G_{i j} = \left \langle \partial_i \psi_0  | \partial_j \psi_0\right \rangle - \left \langle \partial_i \psi_0  | \psi_0\right \rangle\left \langle \psi_0  | \partial_j \psi_0\right \rangle \;.
\end{equation}

This tensor therefore signifies the distance between the two quantum states \cite{KOLODRUBETZ20171}. The Fubini-Study metric is the real part of this tensor, $g_{i j}(\theta) = \mbox{Re}[G_{i j}(\theta)]$. We can view the Fubini-Study metric as a distance measure between the wave functions, or transition probability between the states \cite{Brody_2001}. 

Consequently, $G_{ij}$ can be calculated on quantum hardware \cite{Stokes_2020}. We consider a variational circuit where each layer $l$ is parametrised by $\theta_l$ and includes gates $U(\theta_l)$. These gates U and their functional relation to the Hermitian generator matrix V are described in Section \ref{Sec:MC} and Eq. \ref{eq:exp}, resulting in the relations
\begin{eqnarray}
\label{eq:partial_i}
\partial_i U_l(\theta_l) = -iV_iU_l(\theta_l) \; , \nonumber \\
\partial_j U_l(\theta_l) = -iV_jU_l(\theta_l)\; ,
\end{eqnarray}
where $V_i$ and $V_j$ are Hermitian generator matrices. From Eq.~(\ref{eq:partial_i}) we can find 
\begin{equation}
\label{eq:G_V_i}
\left \langle \partial_i\psi _\theta |\partial_j \psi _\theta \right \rangle =  \left \langle \psi _l | V_iV_j | \psi _l\right \rangle \; ,
\end{equation}
\begin{equation}
\label{eq:G_V_ii}
i\left \langle \psi _\theta |\partial_j \psi _\theta \right \rangle =  \left \langle \psi _l | V_j | \psi _l\right \rangle \; .
\end{equation}

By considering both \ref{eq:G_V_i} and \ref{eq:G_V_ii} a representation of the Quantum Geometric Tensor can be formed for a block of parameters that exist in layer $l$
\begin{equation}
\label{eq:G_K}
G_{ij}^{l} =  \left \langle \psi_l | V_i V_j|\psi_l\right \rangle - \left \langle \psi_l | V_i|\psi_l\right \rangle\left \langle \psi_l | V_j|\psi_l\right \rangle \; .
\end{equation}

The quantum states $\psi_l$ can be determined experimentally from the variational quantum classifier. Importantly, this approximation of the QGT also allows to find the Fubini-Study metric by taking the real part, such that $g_{i j}^{l} = \mbox{Re}[G_{i j}^{l}]$. To calculate the inverse, we use the Moore-Penrose pseudo inverse
\begin{equation}
\label{eq:g+}
g^+ = (g^Tg)^{-1}g^T \;.
\end{equation}
This method allows to finding an inverse matrix even if the matrix cannot be inverted, as shown in Eq \ref{eq:g+}. In cases where the matrix is invertible the matrix pseudo inverse and matrix inverse are identical.

\bibliographystyle{inspire}
\bibliography{references}

\end{document}